\documentclass[sn-mathphys-num]{sn-jnl}

\usepackage{graphicx}%
\usepackage{multirow}%
\usepackage{amsmath,amssymb,amsfonts}%
\usepackage{amsthm}%
\usepackage{mathrsfs}%
\usepackage[title]{appendix}%
\usepackage{xcolor}%
\usepackage{textcomp}%
\usepackage{manyfoot}%
\usepackage{booktabs}%
\usepackage{algorithm}%
\usepackage{algorithmicx}%
\usepackage{algpseudocode}%
\usepackage{listings}%

\usepackage[T1]{fontenc}
\usepackage{lmodern}
\usepackage{fix-cm}
\usepackage{caption}
\usepackage{subcaption}

\usepackage[normalem]{ulem} 

\begin{document}

\title[Design and evaluation of a serious game in virtual reality to increase empathy towards students with phonological dyslexia]{Design and evaluation of a serious game in virtual reality to increase empathy towards students with phonological dyslexia}


\author*[1,4]{\fnm{José Manuel} \sur{Alcalde-Llergo}}\email{jose.alcalde@unitus.it}

\author[1]{\fnm{Andrea} \sur{Zingoni}}\email{andrea.zingoni@unitus.it}

\author[2]{\fnm{Pilar} \sur{Aparicio-Martínez}}\email{n32apmap@uco.es}

\author[3]{\fnm{Sara} \sur{Pinzi}}\email{qf1pinps@uco.es}

\author*[4]{\fnm{Enrique} \sur{Yeguas-Bolívar}}\email{eyeguas@uco.es}

\affil[1]{\orgdiv{Department of Economics, Engineering, Business and Society}, \orgname{University of Tuscia}, \orgaddress{ \city{Viterbo}, \postcode{01100}, \country{Italy}}}

\affil[2]{\orgdiv{Department of Nursing, Physiotherapy and Pharmacology}, \orgname{University of Córdoba}, \orgaddress{ \city{Córdoba}, \postcode{14004}, \country{Spain}}}

\affil[3]{\orgdiv{Vicerectorate for equality and inclusion}, \orgname{University of Córdoba}, \orgaddress{ \city{Córdoba}, \postcode{14071}, \country{Spain}}}

\affil[4]{\orgdiv{Department of Computing and Numerical Analysis}, \orgname{University of Córdoba}, \orgaddress{ \city{Córdoba}, \postcode{14014}, \country{Spain}}}

\abstract{Dyslexia is a neurodevelopmental disorder estimated to strike approximately 5-10\% of the population. In particular, phonological dyslexia causes problems in connecting the sounds of words with their written forms. Consequently, affected individuals may encounter issues such as slow reading speed, inaccurate reading, and difficulty decoding unfamiliar words. To address these complexities, the use of compensatory tools and strategies is essential to ensure equitable opportunities for dyslexic students. However, the general underestimation of the issue and lack of awareness regarding the significance of support methodologies pose significant obstacles. One of the ways to enhance consciousness towards a certain issue is by stimulating empathy with whom is affected by it. In light of this, this study introduces a serious game in virtual reality, targeted at educators, students, and, in general, at the non-dyslexic community. The game seeks to enhance understanding of the challenges that individuals with dyslexia experience daily, highlighting the relevance of supportive measures. This approach encourages players to empathize with the struggles of dyslexic individuals and to learn firsthand the importance of supportive methodologies. The final version of the experience was tested by 101 participants and evaluated through a specific collection of questionnaires validated in the literature. The results show that using the proposed virtual reality tool to promote empathy for individuals with phonological dyslexia is highly effective, leading to an average 20\% increase in participants' empathy after playing the game.}

\keywords{Virtual reality, Inclusive methodologies, Dyslexia, Specific learning disorders, Empathy, University students}

\maketitle

\section{Introduction}\label{sec1}

Since the initial research in~\cite{Morgan1896}, the concept of dyslexia has been changing over the centuries~\citep{Artiles2020}. Dyslexia is defined as a neurobiological-specific learning disorder (SLD) whose impact affects cognitive and social development~\citep{Butterfuss2018,Farah2021}. It is estimated that between 5\% and 10\% of the global population suffers from it~\citep{costantini_psychosocial_2020}, although recent studies indicate that it is underdiagnosed and could affect up to 17\%~\citep{Farah2021,Gabrieli2009}. Since dyslexia is a learning disorder, it is directly linked to language and reading issues, resulting in abnormalities of executive functions~\citep{Farah2021,Gabrieli2009}. Individuals with dyslexia show irregularities in behavioral and neurobiological domains, present in childhood and continuing through the individual's life. These issues translate into specific disorders regarding reading, memorization~\citep{kohli_specific_2018}, social impairment, and neuropsychological issues, such as low self-esteem, social anxiety or depression~\citep{Livingston2018,Christopher2012}.

For students affected by dyslexia, higher education can present significant challenges, especially if no compensatory strategies have been put into practice at the beginning of their learning journey. This increases their insecurities and worsens their physiological wellbeing~\citep{Jordan2014}. In addition to the academic requirements of higher education, dyslexic students may also face a lack of understanding and support from their peers and educators~\citep{Nevill2022}. Recent studies indicated the lack of awareness and understanding of students with dyslexia, turning the teachers into one of the barriers in the learning process~\citep{Rohmer2022}.

Moreover, notwithstanding the increasing acknowledgment of dyslexia as a distinct learning challenge, a substantial disparity persists in the assistance and resources available to dyslexic students at the higher education level~\citep{Gregory2022}. In a prior study by \cite{zingoni_investigating_2021}, the average annual accumulation of university credits within the European Credit Transfer and Accumulation System (ECTS) for students with and without dyslexia in Italy was investigated. The overall findings suggest that students with dyslexia, on average, accumulate six ECTS less per year, roughly equivalent to missing one exam annually compared to their non-dyslexic peers. This highlights the considerable challenges faced by dyslexic students, emphasizing the need for compensative measures to ensure equal opportunities.

Nowadays, research has identified different profiles or sub-types of dyslexia, which can aid in better understanding the phenomenon and addressing individuals' specific challenges. These are (i) phonological dyslexia, (ii) surface dyslexia and (iii) deep dyslexia~\citep{lukov_dissociations_2015}. Phonological dyslexia is characterized by difficulty in reading {both unfamiliar real words (i.e., words not previously encountered by the reader) and non-words (i.e., meaningless letter strings), while maintaining relatively strong performance with familiar words}~\citep{tree_phonological_2006,Marshall1987,Basso2008}. Individuals with phonological dyslexia commonly struggle with connecting the sounds of words to their written representations based on patterns of errors in the subword-level reading route, which can hinder their capacity to recognize and recall new vocabulary~\citep{Marshall1987,Luzzatti2006}. Consequently, this condition can manifest as reading challenges, including reduced reading speed, inaccuracies, and difficulty deciphering unfamiliar words~\citep{rapcsak_phonological_2009}. The underlying cause of phonological dyslexia is believed to derive from some brain variations, particularly in regions responsible for phonological processing~\citep{alexander_lesion_1992}, which are crucial for how language is processed. Nevertheless, with the right support and interventions, individuals dealing with phonological dyslexia can enhance their reading and writing skills, {mitigating challenges related to new vocabulary acquisition, reading fluency, and comprehension, that may otherwise hinder academic and professional development.}~\citep{Maculada2023}.{While direct language-based interventions are fundamental for students with phonological dyslexia, promoting their inclusion alongside classmates and teachers serves as a complementary support that fosters understanding and reduces stigma}~\citep{Hamilton2023}.

Students with dyslexia are likely to face misunderstanding or stigma from their peers and educators; therefore, fostering empathy among the people surrounding them stands as a fundamental objective. Double empathy theorem frames the reciprocal deficits in understanding that may arise when individuals have varying norms and expectations of each other, leading to potential miscommunication~\citep{Hamilton2023,Luzzatti2006}. Demonstrating empathy towards students with dyslexia can help to create a supportive and inclusive learning environment where they feel understood, validated, and respected~\citep{sako_emotional_2016}. Such an environment can significantly enhance their confidence and motivation to learn and, ultimately, contribute to their academic and professional achievements. 

One of the emerging technologies that has gained popularity in promoting empathy over the last decade is virtual reality (VR). It offers three basic possibilities: simulating interactions between different groups, providing a bystander's view of discriminatory actions, or enabling participants to experience discriminatory behaviors from the perspective of an outgroup member or a victim~\citep{Banakou2020,Seinfeld2022}. In particular, the third aspect offers an interesting possibility related to the above-mentioned lack of empathy issue. 

To this end, we implemented ``In the shoes of dyslexic students: The Magic Potion'' (hereinafter simplified as ``The Magic Potion''), a serious VR game designed to enhance comprehension of dyslexia and raise awareness about the challenges it can pose. This document outlines the application's design and the methodology used to validate its functionality through literature-validated questionnaires. It also presents the results obtained from testing the application with over 100 participants. Our aim is to encourage the development of a supportive environment that facilitates academic success for dyslexic students. This work is part of VRAIlexia project (\href{https://vrailexia.eu}{https://vrailexia.eu}), an initiative designed to provide support for university students with dyslexia. The main objective of the project is to mitigate the challenges arising from dyslexia among students in higher education, with the goal of reducing the incidence of university dropouts and facilitating access to degree programs for individuals with dyslexia.

\section{Related works}\label{sec:Related}

As the potential of VR applications continues to be explored across diverse domains, its transformative impact on fostering inclusivity, nurturing empathy, and addressing specific learning challenges, such as dyslexia, has emerged as a compelling area of research.
VR possesses significant potential in cultivating transformative experiences that can propel inclusivity. This capability stems from its capacity to generate interactive and immersive environments capable of replicating everyday scenarios. These scenarios can primarily be applied in two ways within the realm of inclusion: replicating environments inaccessible to individuals from disadvantaged backgrounds, and simulating scenarios that allow awareness about the challenges faced by marginalized groups to be raised.

Firstly, the replication of environments where an individual might not feel physically comfortable allows people to adapt to the surroundings and events in a simulated manner before transitioning to reality. The gradual approach to the uncomfortable environment gets them used to it and stimulates the capability to face it. An illustrative case study of this is demonstrated in \cite{de_luca_virtual_2023}. This project is a social inclusion initiative that utilizes VR and spatial augmented reality technologies to create safe spaces for collaborative activities based on art therapy techniques and new technologies. The primary goal of the project is to acknowledge and amplify individual distinctions while concurrently encouraging collaboration and ongoing interaction among participants, educators, and experts in cultural heritage and technology. 
Another application of this kind involves the inclusion of individuals with disabilities in the labor market~\citep{piovesan_virtual_2013}. In this case, authors leverage immersive environments to enhance employment opportunities for people with disabilities. The system aims to foster workforce inclusion by simulating work-related tasks and facilitating skill development through engaging and interactive experiences. It is worth mentioning that such VR applications can also be extended to specific target groups, such as children with autism. In~\cite{didehbani_virtual_2016}, researchers conducted a case study to explore the effects of ``Virtual Reality Social Cognition Training'', aimed at improving the social skills of children diagnosed with autism spectrum disorder (ASD). During the study, the performance of 30 children diagnosed with ASD was assessed across various domains. This assessment involved exposing them to diverse scenarios, such as collaborative classroom projects and ordering food in the school cafeteria. The study's results underscore the potential of employing a VR platform as a promising therapeutic intervention for gaining deeper insights into the social challenges commonly experienced by individuals with ASD. Notably, the study revealed significant improvements in areas such as emotion recognition, social attribution, and the executive function of analogical reasoning.

The second way to make inclusion benefit from VR is to reproduce issues encountered daily by individuals from a disadvantaged social group. Through transporting users to virtual environments, VR enables the exploration of diverse perspectives, cultures, and identities, potentially eliciting empathy. Numerous studies, including one conducted in~\cite{stilinovic_facilitating_2017}, have demonstrated the benefits of VR in achieving this objective. In this particular one, participants were selected randomly to watch a documentary presented in a VR format, which depicts the life of a young girl residing in a refugee camp. The findings from this research revealed that VR has the potential to significantly enhance the experience of empathy among its users, due to the sense of immersion it creates. In addition, major companies are investing in the potential of VR to raise societal awareness about the challenges faced by various vulnerable groups. Meta, for instance, has embarked on its initiative ``VR for Good'', which narrates stories focusing on individuals from underprivileged social groups. Examples of such initiatives include ``We Live Here''~\citep{MetaWeLiveHere}, providing users an insight into the life of a homeless person; ``Home After War''~\citep{MetaHomeWar}, depicting the story of a refugee returning to a war-torn homeland facing the fears of an unsafe environment post-war; or ``The Hidden''~\citep{MetaTheHidden}, shedding light on the life of an enslaved family, aimed at raising awareness about the persistence of slavery in today's world.

Nevertheless, the potential of VR to cultivate empathy extends beyond these instances. Apart from recreating real scenarios, VR also allows to simulate different symptoms or disabilities experienced by disadvantaged individuals. An exemplary case illustrating this potential is found in ``Notes on Blindness: Into Darkness''~\citep{arteNotesOnBlindness}, where users are engaged in an experience that aims to place them in the shoes of someone who is gradually losing vision until becoming completely blind. The experience is narrated through audio recordings by this individual, describing how he visualized the world through sound.
Another notable application involving the replication of symptoms experienced by individuals with disabilities through VR was developed in~\cite{hoter_effects_2022}. The primary aim of this research was to assess the impact of a simulation of the obstacles faced by students in wheelchairs while performing their everyday tasks. The study's findings demonstrated that this simulation experience effectively transformed participants' attitudes toward individuals with disabilities in the real world, motivating them to become advocates for positive change. This study contributes valuable insights to the growing body of research highlighting the significance of VR simulations in enhancing empathy towards others.

Similarly, VR can be employed for the inclusion of individuals with dyslexia. Its application as a tool to address the challenges posed by this SLD in children has expanded in recent years, leading to a rise in research endeavors in this domain. Notably, studies such as the one conducted in~\cite{pedroli_psychometric_2017} have indicated that VR can have a positive impact on the memory and skills of individuals with dyslexia. In this particular study, participants engaged in various assessments within a virtual classroom setting, where they were required to respond to tasks presented on a blackboard. While the results did not demonstrate significant enhancements in reading proficiency, they revealed a noticeable improvement in participants' attention levels. This heightened attention could potentially lead to long-term improvements in other challenges associated with dyslexia, such as reducing the time required to read low-frequency long words.

Furthermore, another typical category of applications designed to support individuals with dyslexia includes games aimed at enhancing users' reading abilities~\citep{Ostiz-Blanco2021,Brennan2022}. There are also games intended to detect whether children have dyslexia while they practice various reading exercises, such as in the case of ``Dytective''~\citep{Rello2017}. This kind of applications have also been extrapolated to the realm of VR. A clear example of such kind of initiatives is the one developed by the European project FORDYSVAR~\citep{rodriguez_cano_tecnologias_2021}. This work's primary objective is to provide a technology-based approach to facilitate the learning process for individuals with dyslexia, with a specific emphasis on leveraging virtual and augmented reality. The project is particularly geared towards assisting dyslexic children aged between 10 and 16 years. As part of its contributions, the project team has developed supportive software tailored to children with dyslexia. This software, presented in~\cite{rodriguez-cano_design_2021}, offers an engaging alternative for addressing various learning challenges associated with dyslexia. It takes the form of a VR video game designed for the Oculus Quest platform, offering an enjoyable and interactive means for dyslexic children to work on alleviating the impact of dyslexia on their learning experiences.

It is noteworthy that the majority of studies pertaining to the application of VR in the context of dyslexia primarily focus on individuals at levels of education below higher education. An exception to this is represented by VRAIlexia project, which is specifically designed with the objective of assisting students in higher education by harnessing VR and artificial intelligence technologies. Among its works, the study by \cite{zingoni_ML_2023} demonstrates how to provide personalized support methodologies to dyslexic students through the application of various machine learning techniques. Moreover, within the project, various VR applications were designed and developed. These include one for the administration of specific psychometric tests to students with dyslexia~\citep{yeguas-bolivar_determining_2022}, in addition to the serious game presented within this document.

\section{Modeling phonological dyslexia for empathy}\label{Overview}

The different stages to model phonological dyslexia to stimulate empathy in the users were accurately planned, starting from the formulation of the VR experience framework. In particular, first, the simulation of reading difficulties inherent in individuals with dyslexia was delineated. Subsequently, the establishment of the experiential context, which encompasses the spatial configuration, objects, and main characters with which the user will interact within the game, is outlined in detail.

The game was developed using the Unity game engine~\citep{unity_2014} to be deployed on the Meta Quest 2 {\& 3~\citep{Meta} platforms}, where its functionality will be validated. Its development followed Scrum methodology~\citep{Schwaber2020}, involving iterative development and testing of the various game versions until reaching the final application.

\subsection{Simulating reading difficulties}
\begin{figure}[ht]
    \centering
    \includegraphics[width=0.5\textwidth]{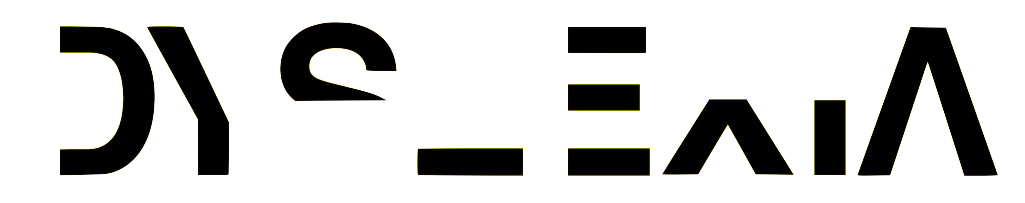}
    \caption{The word ``DYSLEXIA'' using Britton's font.}
    \label{fig:britton}
\end{figure}

The methodology proposed in this study involves immersing participants, including teachers or peers of dyslexic students, into a virtual world to execute a specific task. This task entails reading a recipe book and precisely integrating the ingredients in the specified sequence and amount, incorporating an additional reading challenge to simulate phonological dyslexia. To replicate the reading difficulties experienced by individuals with dyslexia, the text in the book and ingredient labels are presented in Britton's Dyslexia font~\citep{britton}. This font was designed by eliminating certain parts of letters, thereby intensifying the reading challenge, even for people not affected by dyslexia disorder. Figure~\ref{fig:britton} shows how the word ``Dyslexia'' is written using this font.

\subsection{Game context}

The serious game has been designed as follows. Upon entering the game, players find themselves within the confines of a virtual ``magic castle'', positioned in front of a closed door. The design of the fantasy virtual scenarios served a dual purpose. Firstly, it capitalizes the widespread popularity of fantasy themes, enhancing the game's overall appeal. Secondly, this design choice set the stage for a meaningful context for the task at hand. The utilization of fictional virtual environments is a prevalent practice, aiming to immerse players in an alternate reality. This approach ensures engagement with the presented task while retaining control over the overall gaming experience~\citep{MetaTheKey}. Then, when players go through the door they meet their ``potions teacher'', who is an avatar that explains them that they must prepare a potion to save their friend, named Sam. The teacher indicates that the potion can be created by adding different ingredients, but by strictly respecting the order to pour them into the pot. The order is written within a book they can find on a table when the game begins. Conversely, when players fail to prepare the potion within the established time limit, they will observe a deterioration in Sam's condition, accompanied by the teacher's admonishment for the inadequate completion of the task.

\subsection{Virtual rooms}

\begin{figure}[ht]
    \centering
    \includegraphics[width=0.9\textwidth]{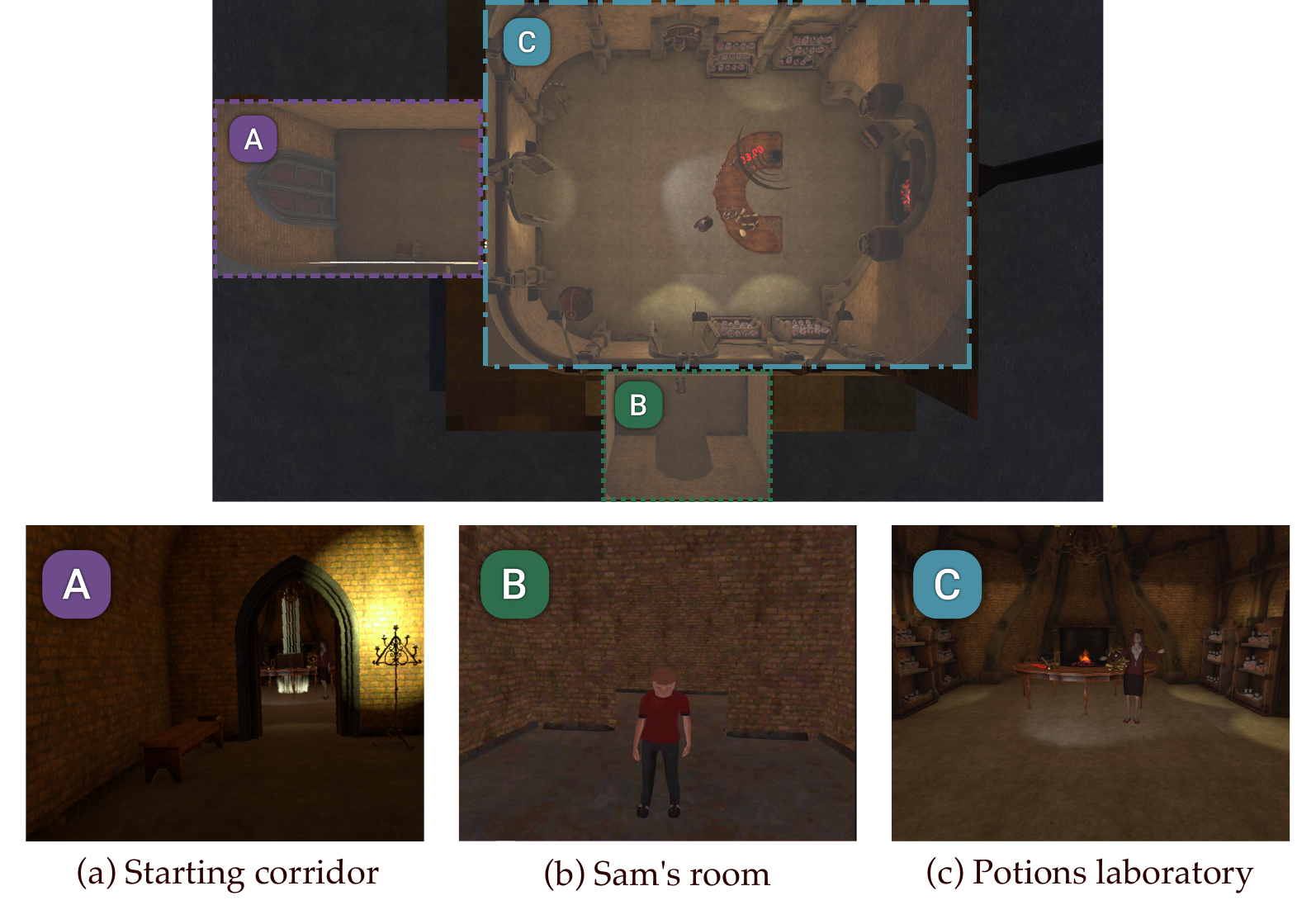}
     
        \caption{{Chambers from the virtual castle: (a) corridor to access the lab, (b) Sam's room and (c) potions laboratory.}}
        \label{fig:rooms}
\end{figure}

The virtual environment is characterized by a dark and dim atmosphere, intentionally crafted to evoke feelings of desolation, fear, and helplessness~\citep{Jordan2014, Livingston2018} in the face of the impending challenge. Within the virtual castle, participants will be granted admittance to three distinct rooms{, organized as shown in Figure~\ref{fig:rooms}}. The initial of these enclosures is the starting point of the exploration, where players can familiarize themselves with the operational features of the controller buttons (Figure~\ref{fig:rooms} (a)). In addition, they will find a brief explanation about the effects of phonological dyslexia. Second room is Sam's room. It is an empty room where players can see their friend and his current state as shown in Figure~\ref{fig:rooms} (b). Finally, the third room is where the game takes place. It is a potions laboratory where players can find their teacher, the table where the potion instruction book is placed and all the ingredients needed to brew the potion to help Sam, which are laid out on shelves. A first view of the lab is shown in Figure~\ref{fig:rooms} (c).

\subsection{Main characters and items}

\begin{figure}
        \centering
        \begin{subfigure}[b]{0.3\textwidth}
             \centering
             \includegraphics[width=\textwidth]{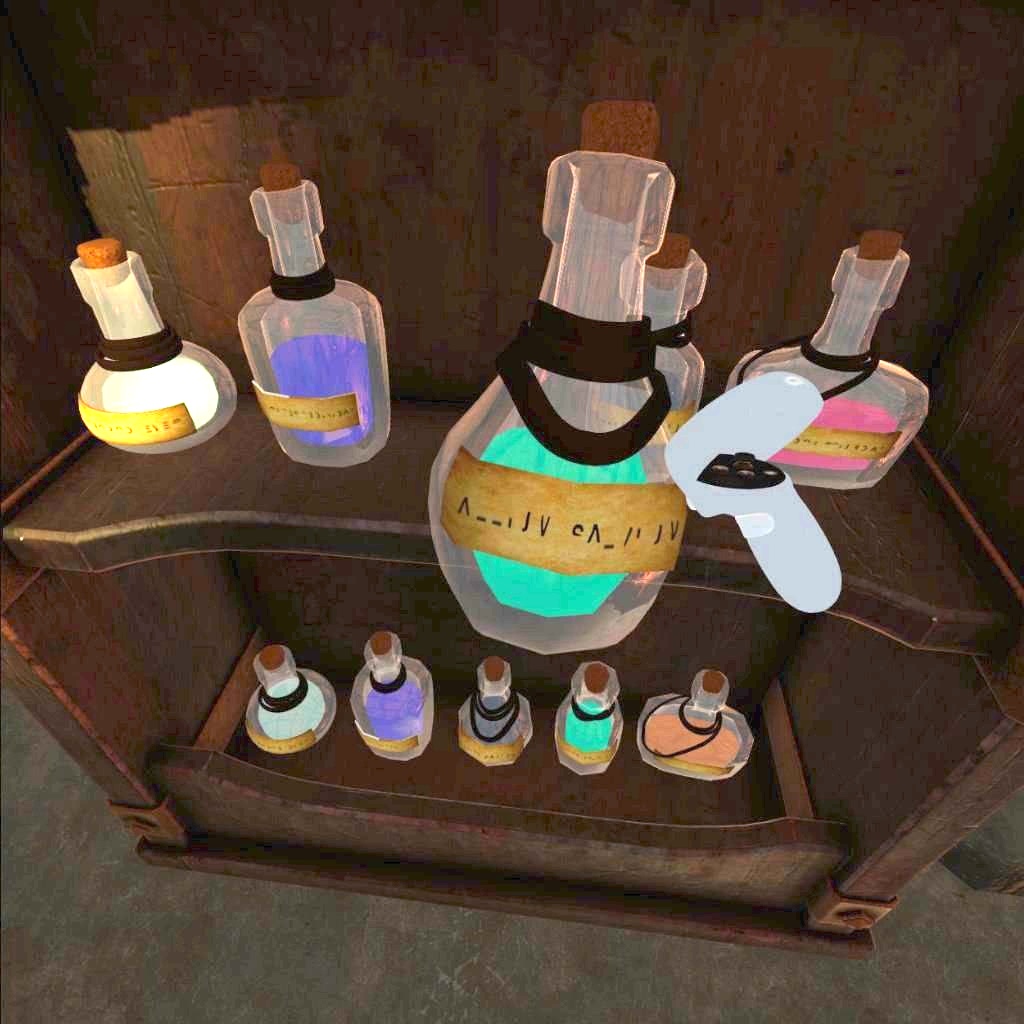}
             \caption{Ingredients.}
             
             \label{subfig:shelve}
        \end{subfigure}
        \begin{subfigure}[b]{0.3\textwidth}
             \centering
             \includegraphics[width=\textwidth]{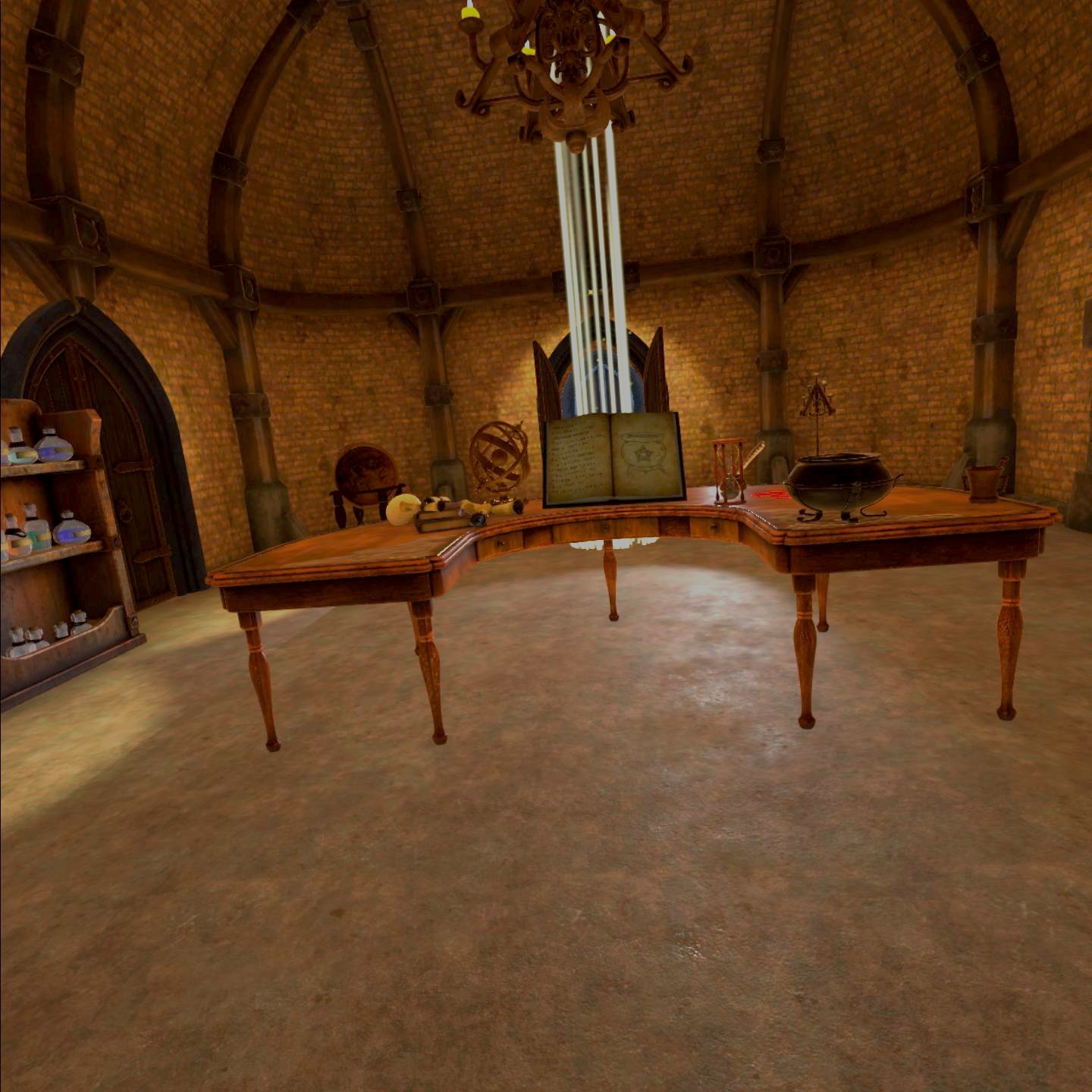}
             \caption{Irregular table.}
             \label{subfig:table}
        \end{subfigure}
        \begin{subfigure}[b]{0.3\textwidth}
             \centering
             \includegraphics[width=\textwidth]{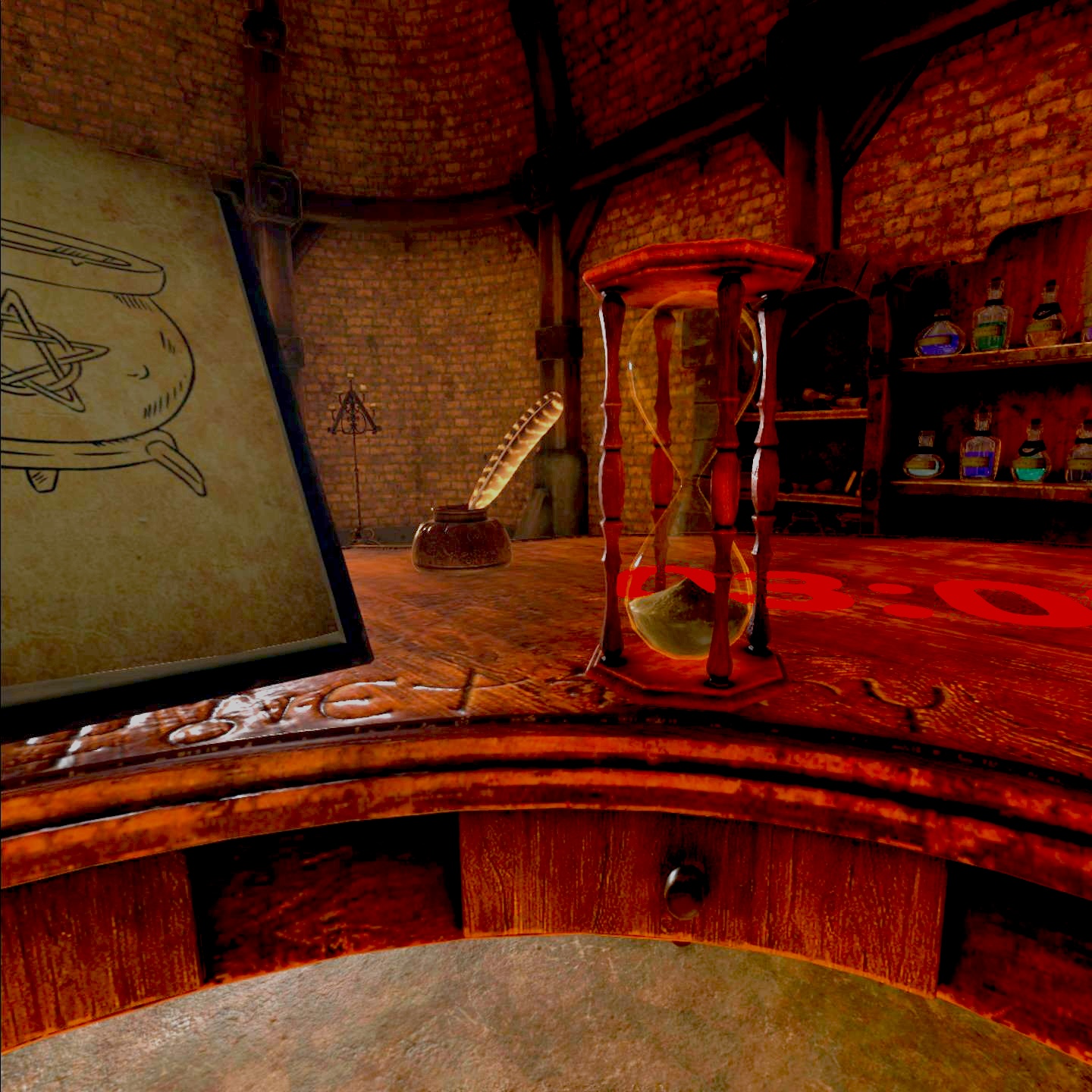}
             \caption{Hourglass.}
             \label{subfig:hourglass}
        \end{subfigure}
        \begin{subfigure}[b]{0.3\textwidth}
             \centering
             \includegraphics[width=\textwidth]{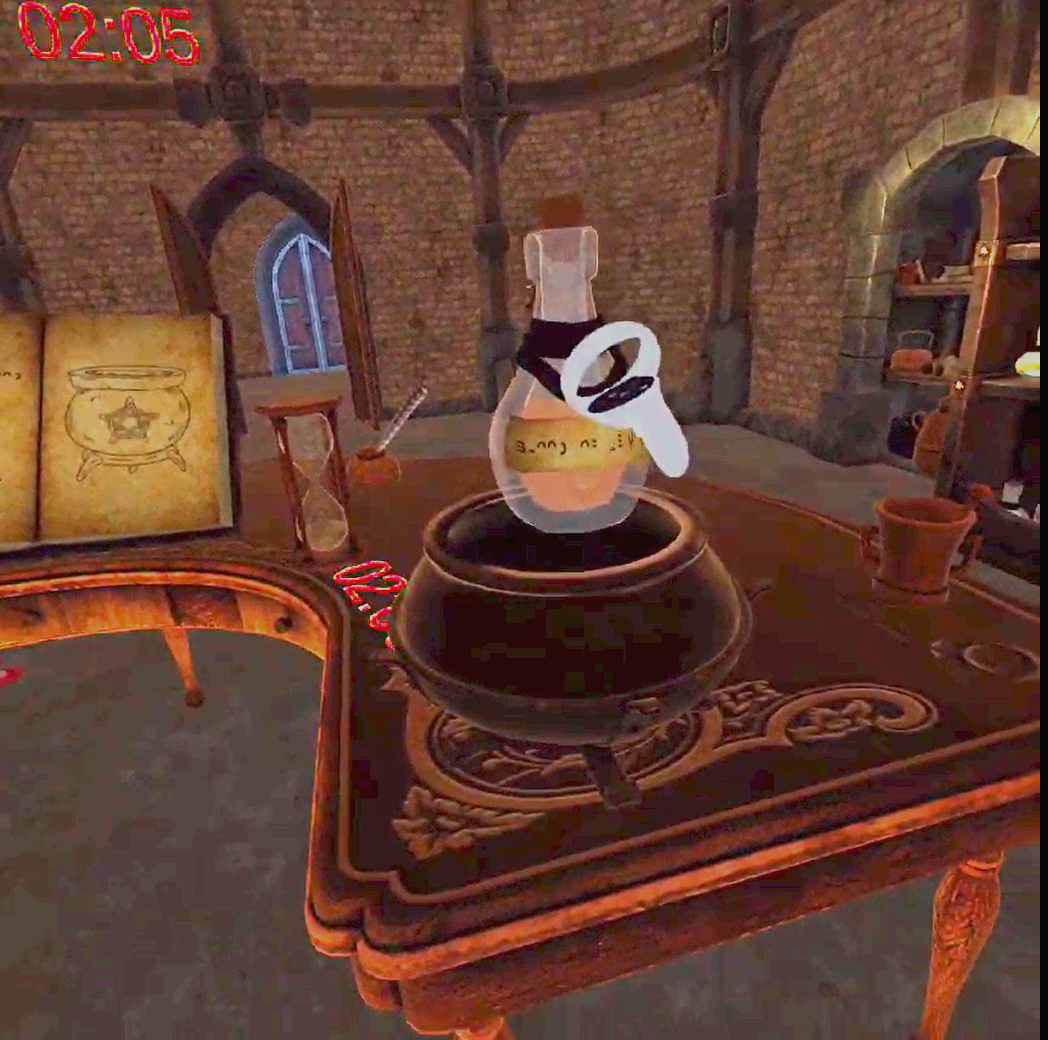}
             \caption{Pot.}
             \label{subfig:pot}
        \end{subfigure}
        \begin{subfigure}[b]{0.3\textwidth}
             \centering
             \includegraphics[width=\textwidth]{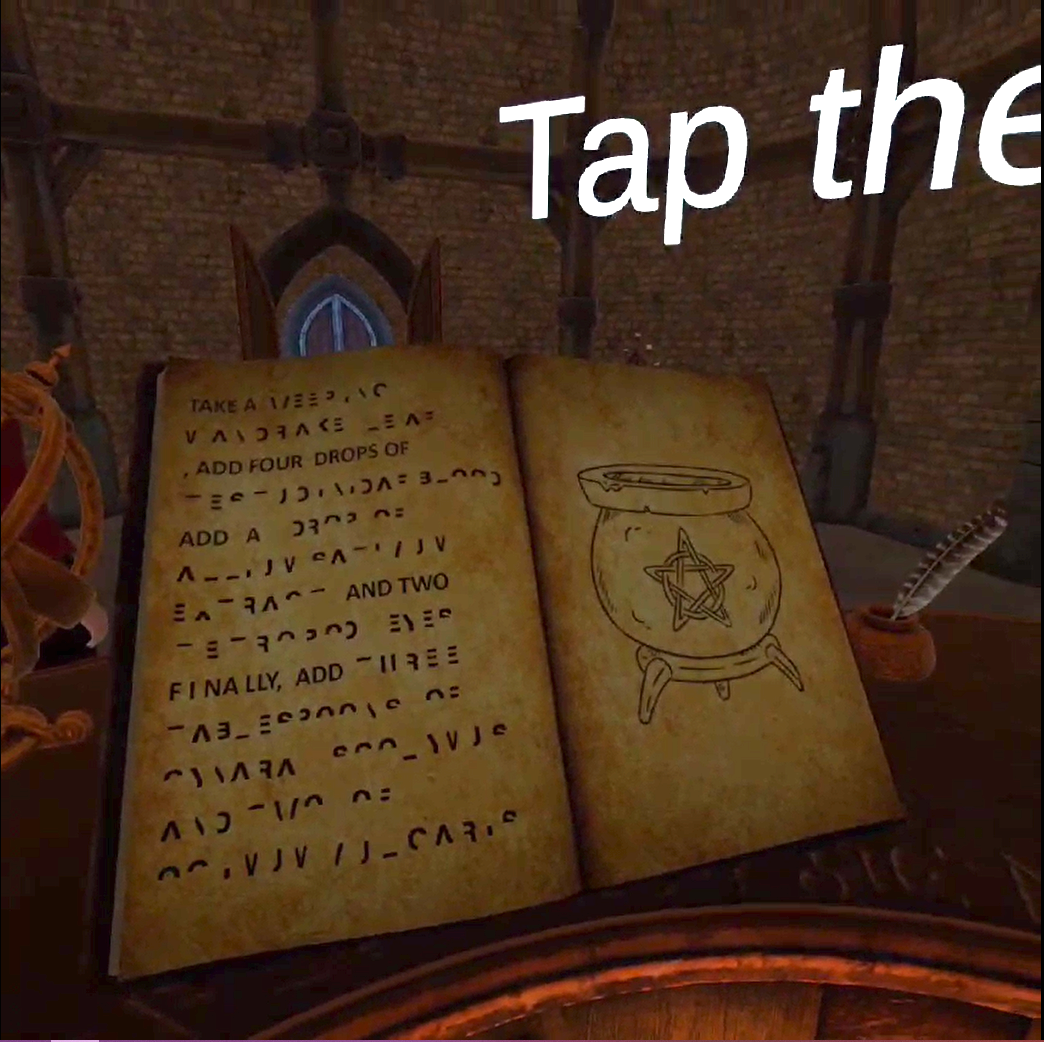}
             \caption{Recipe book.}
             \label{subfig:book}
        \end{subfigure}
        \begin{subfigure}[b]{0.3\textwidth}
             \centering
             \includegraphics[width=\textwidth]{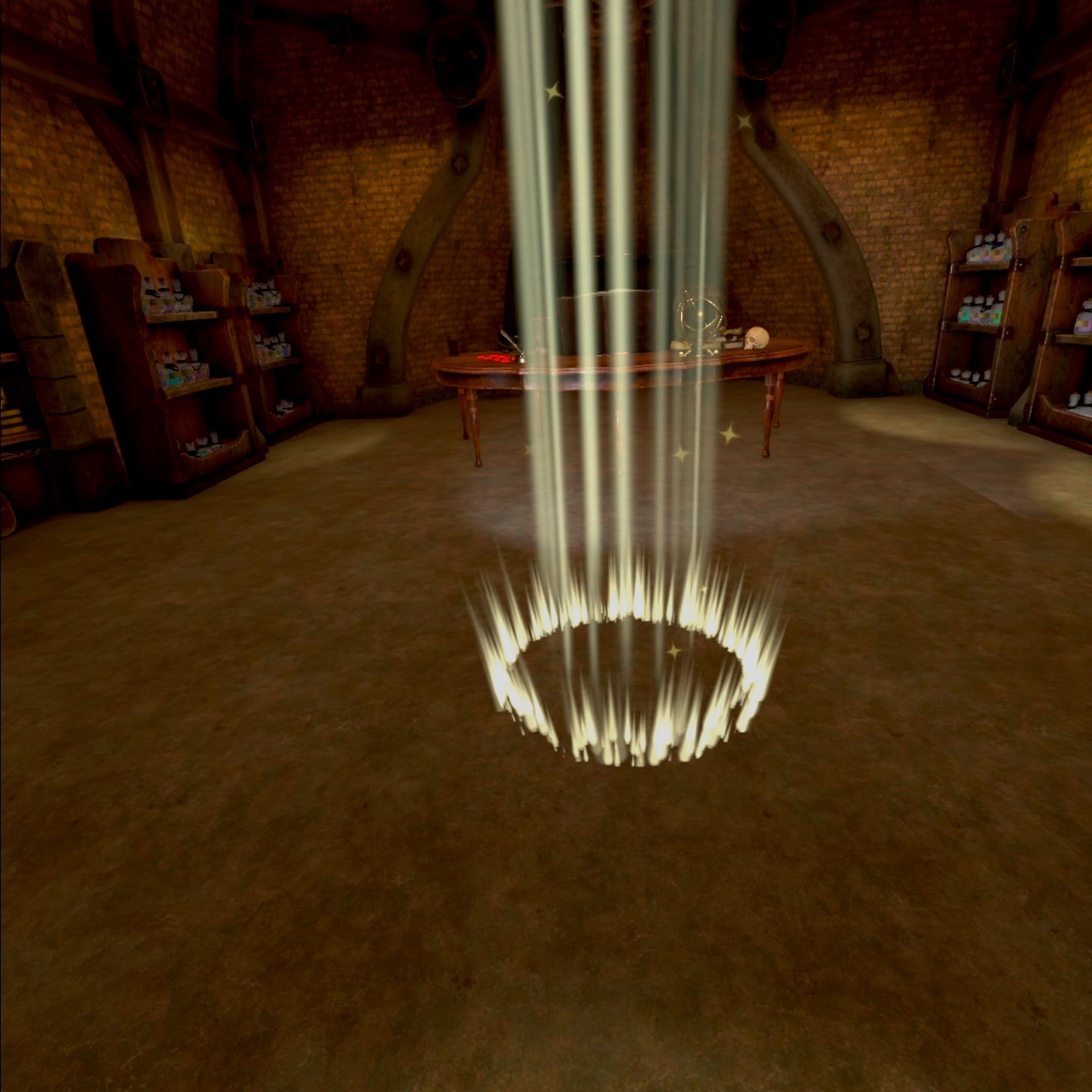}
             \caption{Beacon.}
             \label{subfig:beacon}
        \end{subfigure}
        \caption{Main items from ``The Magic Potion''.}
        \label{fig:main_items}
    \end{figure}

Throughout the VR experience, participants are required to interact with a diverse range of items and characters that have been specifically designed to guide and enhance their level of immersion within the simulated environment. A list of these characters and objects is exposed below, and their representation in the virtual world is shown in Figure~\ref{fig:main_items}.

\begin{itemize}
    \item \textbf{Teacher}: this non-player character (NPC) has the role of potions teacher. The avatar explains to the players the rules of the game belonging to the different levels and what they have to do to create the correct potion and win the game. In addition, this NPC takes on the role of a strict teacher who yells at the students when they fail in their tasks. In this way, the player put himself in the place of dyslexic students when they are not valued for their effort by those teachers~\citep{Nevill2022, Rohmer2022}, that only keep performance in mind.
    \item \textbf{Sam}: represents the player's friend in the game. The player must prepare a potion correctly and within a limited time frame in order to save this NPC. Player's results will be directly reflected in Sam conditions. If the correct potion is created, Sam will be happy and will celebrate, but if users fail during the different levels, they will see Sam suffer.
    \item \textbf{Ingredients and shelves}: around the potions laboratory the player finds several shelves with different ingredients to make potions. These ingredients are carefully arranged in a particular order on each shelf and stored within flasks of varying shapes and colors, each labeled with the corresponding ingredient's name. Notably, the labeling is composed utilizing Britton's font, trying to replicate the reading difficulties experienced by a dyslexic person.
    \item \textbf{Table}: in the center of the potion laboratory lies an irregularly-shaped table designed to promote different modes of locomotion in VR (teleportation-based and controller-based). Over the table, there are different key objects for the game, including the hourglass, the recipe book, and the pot, which players may interact with as part of their gameplay experience. Furthermore, the table serves as a practical space for players to place some of the potion ingredients. 
    \item \textbf{Hourglass}: is the item used to start the game. Next to it there is a digital clock with the time available to complete the level. Upon starting the game, the assigned time begins to decrement, with the level concluding once the timer reaches zero. The design of this item, which constantly displays the remaining time, was created to be overwhelming for the player, in a manner similar to how the deadline of an exam can be stressful.
    \item \textbf{Pot}: container in which the potion has to be brewed. Players must pour the correct ingredients into the pot in the correct order to pass the game. The pot will release a heart if the poured ingredient is correct and a purple smoke if it is not, serving as feedback to guide the player towards successful completion of the game.
    \item \textbf{Recipe book}: it is a big book containing the recipe for the potion to be brewed by the player. It denotes the specific type and quantity of each ingredient required and specifies the correct order in which they must be added to the container, in order to ensure the successful completion of the potion. Consistent with the font utilized for ingredient labeling, the recipe book also employs Britton's font.
    \item \textbf{Beacon}: a light beam that guides the player throughout the virtual environment. It indicates specific points within the potions laboratory where the player must be situated to initiate teacher conversations or start the next level.
\end{itemize}

\section{Gameplay design} 

This section presents a detailed overview of the game design. First, an initial explanation of the player's interaction with the virtual environment (which is displayed on the controllers) is given. Then, the compensatory methodologies considered are introduced. After this, the different stages of the game are differentiated. Following this, the flow of the game is illustrated, encompassing the player's journey from the initial entry into the virtual environment to the culmination of the experience. Finally, a detailed exposition of the various levels of difficulty under consideration is presented.

\subsection{Interaction with the virtual environment. }
Once players are within the virtual world, they can perform various interactions with the objects in the environment through movement and the use of controllers. Degrees of freedom (DoF) in virtual environments refer to the number of ways in which a user can move or interact within the virtual space. These include movements along various axes and rotations, allowing for a more immersive and realistic experience. The implemented experience can be conducted statically in 3DoF, with complete movement control via joysticks, or by moving within a defined safe zone with 6DoF. In both cases, the user is allowed two types of movement through the controllers: linear and parabolic. The linear movement is slow and ideal for reaching nearby places and objects with precision. As for the parabolic movement, it involves the player's movement through teleportation to the end of a parabola launched by pointing the controller towards a point in the virtual environment. The latter is useful especially when there is not enough real space to perform the experience.

Regarding interaction with objects, some, such as the hourglass, are activated simply by touching them. It will also be necessary to grab and transport the different ingredients to create the potion through the environment to complete the task. This can be done using both controllers by placing them on the ingredients they wish to grab and pressing the grab button to grasp. When they want to release the object, they simply need to release the button on the controller corresponding to the hand they are using to hold it.

\subsection{Compensatory tools}
The development of the game has required the translation of various compensatory methodologies for dyslexic students into the virtual environment. These tools have been inspired by real-world methodologies~\citep{benedetti_clustering_2022} and adapted for the tasks to be performed during the game. Three support tools have been considered:
 
\begin{itemize}
    \item \textbf{Time extension:} Each time users fail an attempt in the game, they are granted additional time to complete the task. Initially, the available time is set at three minutes, which, upon failing, increases first to five minutes and ultimately, in the game's final phase, to ten minutes. This system simulates the compensatory educational tool of allotting extra time for a student to complete an exam.
    \item \textbf{Word abbreviation:} After failing the first level, some of the ingredient names, both in the recipe book and the potion labels, are simplified to make them easier for the player to recognize and memorize. An example of simplification can be observed in Figure~\ref{fig:shorterwords}, where the term ``Testudinidae'' is replaced with ``Turtle''. This tool simulates the practice of employing simple {and familiar} language both in explaining lessons and in formulating exam questions for dyslexics {to facilitate recognition and reduce cognitive load}.
        \begin{figure}[ht]
            \centering
            \includegraphics[width=0.9\textwidth]{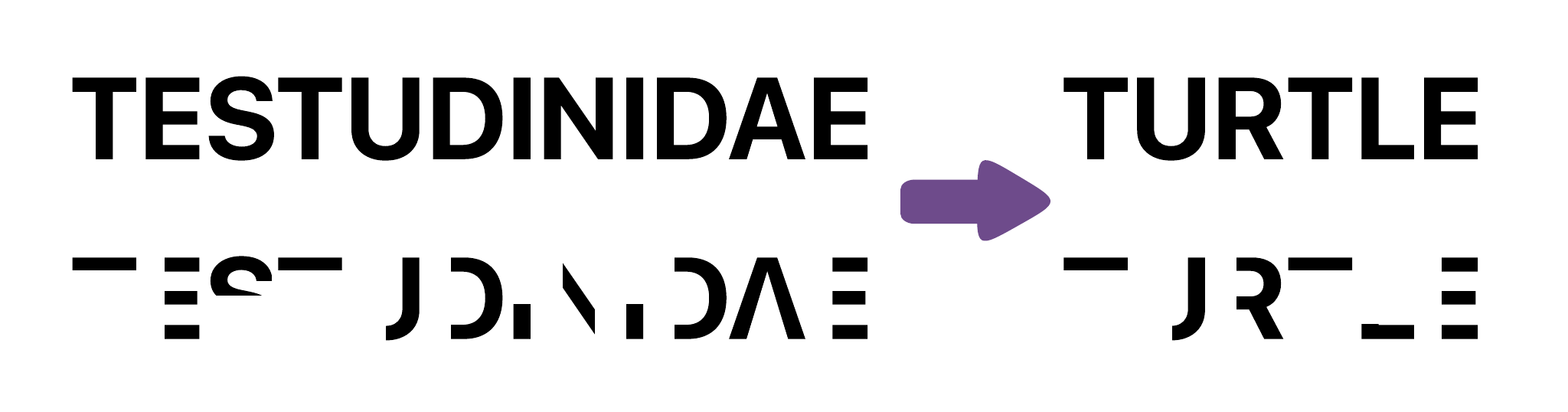}
            \caption{Word shortening compensation tool. Simplification of the term ``Testudinidae''. The words are displayed in both the normal alphabet and their corresponding writing in the game font}
            \label{fig:shorterwords}
        \end{figure}
    \item \textbf{Audio guide:} Upon failing the second level, an auditory aid will be unlocked, so that each time the player touches a word in the recipe book or an ingredient, the system will read the word aloud. This way, players will be able to accurately match the recipe with the different jars of required ingredients. This simulates the use of a real-life audio guide to facilitate the comprehension of complex texts for students with dyslexia, leveraging auditory support to enhance textual understanding.
\end{itemize}

\subsection{Game levels}
The game comprises a total of four distinct phases. The phase zero or the tutorial phase serves as an introductory stage, providing the users with a concise overview of phonological dyslexia, along with an opportunity to familiarize themselves with the game controls. Subsequent to this initial phase, the primary game flow begins, where players are tasked with the preparation of the potion. The primary flow is structured into three levels, each corresponding to a distinct phase of the gameplay experience.

In the first level, the primary objective is to immerse the player in the experience of empathizing with a person challenged by dyslexia. With a limited time of three minutes, the player will confront the challenge of handling lengthy and intricate words, written in a font that is difficult to decode, during the potion-brewing process. Furthermore, to effectively convey a sense of frustration in the event of failure at the end of this phase, the teacher reprimands the player for their inability to accomplish what might be perceived as a straightforward task. Moreover, the timer enhances the sense of anxiety, making players to experience also this issue.

In the second phase, the player is provided with two compensatory tools similar to those that might be provided to a university student with dyslexia when taking an exam. Firstly, an extended time allowance grants the player five minutes to craft the potion. Secondly, the simplification of the terms that describe the ingredients facilitates their identification. In the case of failure, the teacher once more reproves the player, aiming to convey the frustration and sense of incomprehension experienced by a dyslexic student despite earnest efforts to complete any task.

In the concluding phase, the player is once again presented with two additional compensatory tools. Firstly, an extended time limit of up to ten minutes is provided. Secondly, an audio aid feature is introduced, enabling the narration of recipe instructions and ingredient labels upon the player's interaction within the virtual environment. Users can repeat this final phase as many times as needed since, with the use of the compensatory tools provided, they can successfully overcome the experience. There is still negative feedback from the teacher in these cases; however, upon successful completion, corresponding positive feedback is offered, highlighting the benefits for individuals with dyslexia stemming from the aids provided during the game.

\begin{figure}[ht]
    \centering
    \includegraphics[width=0.9\textwidth]{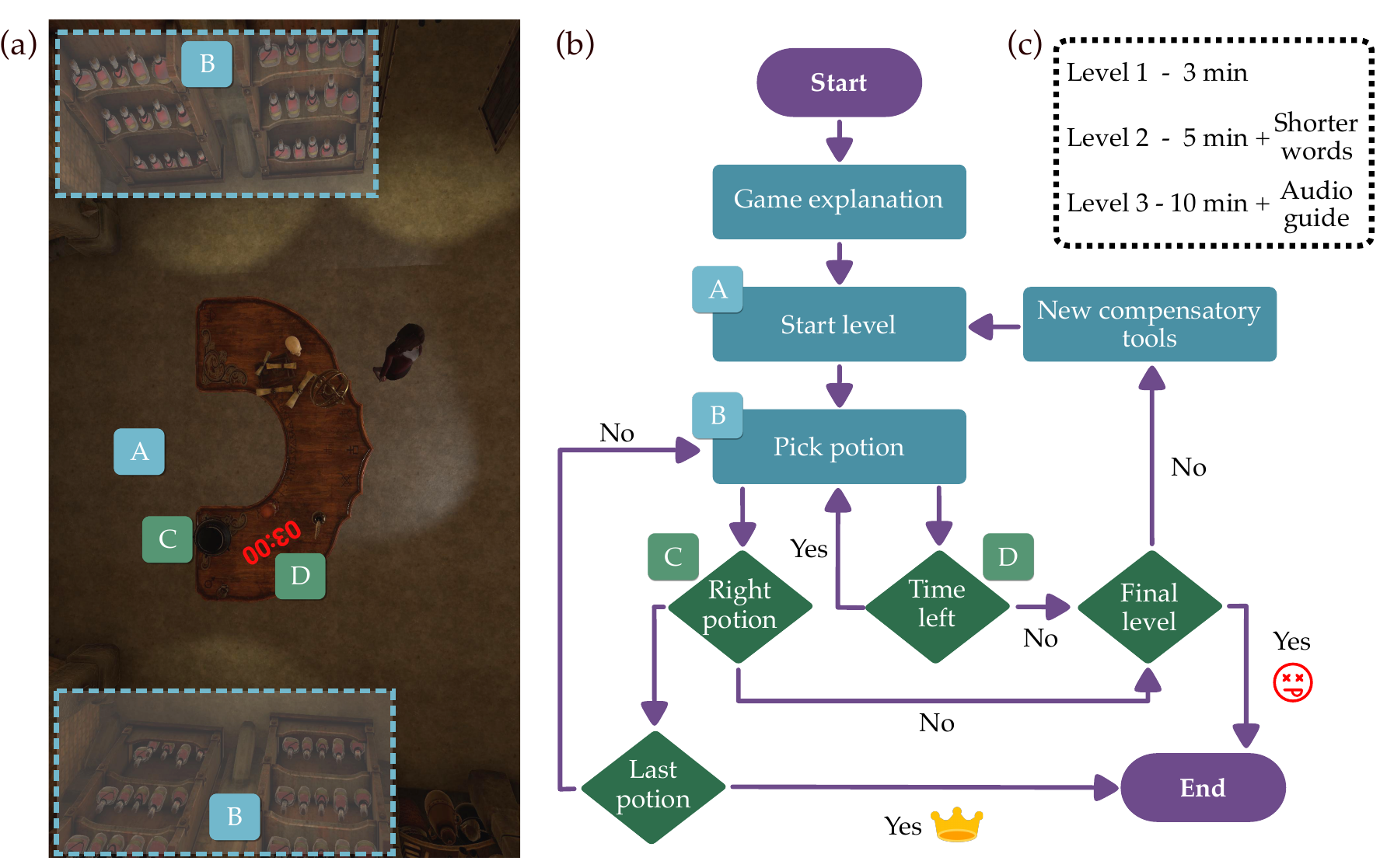}
    \caption{{Overview of ``The Magic Potion'': (a) Main scenario of the game, (b) flow diagram, and (c) compensatory tools included in each level.}}
    \label{fig:flow_diag}
\end{figure}

The depiction of the game's progression throughout the task across the various levels has been presented in Figure~\ref{fig:flow_diag}, which is divided into three subfigures: (a)~displays a view of the potions laboratory within the virtual environment, highlighting key interaction areas labeled as A (Level start point), B (Potion shelves), C (Pot), and D (Timer); (b)~shows the flow diagram guiding the logic progression of the game; and (c)~summarizes the three levels and their associated compensatory tools.

The software is in the process of intellectual property registration and has been developed for Meta Quest 2 {\& 3} in three different languages: English, Spanish and Italian. It is available for download at \href{https://sidequestvr.com/app/37611/}{https://sidequestvr.com/app/37611/}. A Meta Quest 2 private in-game capture of ``The Magic Potion'' can be visited in the following link: \href{https://youtu.be/C80DuTWT3X8}{https://youtu.be/C80DuTWT3X8/}. The evidence that this software has been developed within the VRAIlexia project is available on the website: \href{https://vrailexia.eu/demo/}{https://vrailexia.eu/demo/}.

\subsection{Difficulty levels}
During the game’s development, another scale of difficulty has been set. It consists of two levels of hardness that involve all the three previously mentioned levels of the game.  In the easiest level, there will only be one container per ingredient, and the ingredient's dose is not considered, since just one container per ingredient must be poured into the pot. This simplifies the task by eliminating the need to consider quantities in the recipe. Conversely, in the hardest mode, there will be multiple {labeled} containers per ingredient, {increasing the probability of confusion between visually similar words and amplifying the decoding demands, thereby simulating the bigger reading challenges that individuals with dyslexia often face in more challenging scenarios}. With the inclusion of these two different difficulty levels, the aim is to enable non-dyslexic individuals to observe how dyslexia impacts the performance of tasks of varying complexities~\citep{rapcsak_phonological_2009}.

\section{{Evaluation methodology}}

{This section describes the methodology used to evaluate the effectiveness of the proposed VR experience, both in terms of its impact on empathy and its usability. It includes details on participants, procedure, measurement instruments, and a preliminary testing of the application prior to the final analysis.
}

\subsection{{Participants}}
 A total of 101 non-dyslexic people participated in the full VR experience and answered a series of questionnaires. Among the participants, there were mostly students and professors from the universities of Córdoba (Spain) and Tuscia (Italy), as well as relatives of university students with dyslexia. Moreover, individuals completely unrelated to the environment of dyslexic university students were also allowed to participate. {These participants tested the game during different sessions held during the second semester of the 2022-2023 academic year. The average duration participants needed to finalize the complete experience was approximately 20 minutes.} Complete information collected about participants is shown in Table~\ref{tab:participant_features}.

\begin{table}[h]
\caption{Main features of the participants in the study}\label{tab:participant_features}
\begin{tabular}{l l l l}
\hline
\textbf{Feature} & \textbf{Subgroup}               & \textbf{\textit{\#}}               & 
\textbf{Percentage} \\ \hline
Age     & < 18  & 8 & 7.92                 \\
     & [18, 30]  & 71 & 70.30                 \\
     & > 30  & 22 & 21.78                 \\
Gender & Female  & 57 & 56.44                 \\
       & Male  & 44 & 43.56                 \\
       & Other  & 0 & 0.00                 \\
Participant profile & Dyslexic relative  & 12 & 11.88                 \\
                    & University student  & 47 & 46.53                 \\
                    & University professor  & 22 & 21.78                 \\
                    & Other  & 10 & 9.90                 \\
Experience with VR applications & None  & 52 & 51.49                 \\
        & Only videos  & 8 & 7.92                 \\
        & Yes, but without headset  & 3 & 2.97                 \\
        & Yes, using a headset  & 27 & 26.73                 \\
        & Very familiarized with VR  & 11 & 12.88                \\
Experience with empathy applications & None  & 67 & 66.33                 \\
        & Know some empathy application  & 18 & 17.82                 \\
        & Watch some videos  & 1 & 0.99                 \\
        & Try one empathy application  & 5 & 4.95                 \\
        & Try several empathy application  & 2 & 1.98                \\
\end{tabular}
\end{table}

Given that the study was conducted in a university setting, about 70\% of the participants are between the ages of 18 and 30. It is also worth noting that the gender participation was quite balanced, with 43.56\% male participation compared to 56.54\% female participation.  Regarding the experience with VR, it is evident that the majority of participants had never tried it before or had only done so in the context of a specific VR experience at a technology event. Finally, the participants' experience with empathy applications was even more limited, with over 65\% of them not being familiar with any such application.

\subsection{{Measures}}\label{sec:assessment}

One of the main objectives of this study is to assess the developed VR experience both as a VR application and as a tool for promoting empathy. To achieve this, {a concise set of questions was developed to collect the necessary data for evaluating the user experience and the change in empathy towards dyslexic students}. 
To this end, the methodology defined in~\citep{vr4MedicalTraining} was adapted to collect a suitable set of questionnaires for evaluating the effectiveness of a VR application. {The chosen questionnaires were selected from previous studies in the literature and were specifically prepared and validated to measure the main factors relevant to the} VR game: empathy improvement~\cite{Spreng2009},  sickness~\cite{Kim2018} and presence~\cite{Usoh2000}. 

Regarding the measure of how experience affects user empathy, the Toronto Empathy Questionnaire (TEQ)~\cite{Spreng2009} will be used. {The TEQ was preferred over longer instruments such as the Empathy Quotient~\cite{BaronCohen2004Empathy} due to its brevity and its demonstrated reliability and construct validity in prior studies~\cite{Youseff2014,Innes2022}, making it more suitable for the experience assessment}. However, questions from the original TEQ are formulated to measure empathy without focusing on any specific study group. Therefore, for this work, these questions have been reformulated to target students with dyslexia and make them more suitable for this study. The responses to all the questions are on a 5-point Likert scale, where: Never~=~0; Rarely = 1; Sometimes = 2; Often = 3; Always = 4. It should also be noted that questions 2, 4, 7, 10, 11, 12, 14, and 15 are negative, and the order of the given values must be reversed for calculating the questionnaire score. Final TEQ score is calculated by summing individual scores, which range from 0 to 64.
{The TEQ was administered to the same group of participants before and after the VR experience}. {The second administration was conducted four months later to evaluate the persistence of any changes in empathy. Additionally, to capture the participants’ immediate impressions, qualitative feedback and behavioral observations were also collected directly after the session, as reported in Section \ref{sec:feedback}.}

Another important point is to ensure that the VR experience will not have adverse effects on the participants' health, such as dizziness, nausea, or disorientation. To achieve this, during an initial assessment phase of the application, the participants were subjected to the VRSQ~\cite{Kim2018} {(See Section \ref{sec:sickness})}, a questionnaire based on the SSQ designed by Kennedy et al.~\citep{Kennedy1993} and adapted for modern VR experiences and tools. This questionnaire simplifies the measurement of sickness in virtual environments into just 9 questions classified into two components: oculomotor and disorientation. All items are measured on a 4-point Likert scale, where: Not at all = 0, Slightly = 1, Moderately~=~2, and Very much = 3. The final result of this metric is given by the average of the results obtained for each of the considered components, yielding a final result in the range between 0 and 3.

Additionally, in any virtual environment, it is important to evaluate the level of presence the user feels within that environment. Presence is defined as the experience of being in one place or environment without depending on whether you are physically there or not~\cite{Witmer1998}. To calculate the relative impact of the developed virtual environment, the participants were subjected to the Slater-Usoh-Steed (SUS) questionnaire~\cite{Usoh2000}, one of the most accepted tools in the literature for this task. This questionnaire is composed of 6 items. Each question is answered on a 7-point Likert scale, where 7 represents the highest level of presence. The final SUS score is determined by counting the number of questions rated 6 or 7, but also the average of all the ratings can be considered as a measure of the presence.

Finally, additional questions were included to collect sociodemographic information and to better understand the profile of the users. Then, after distributing the questionnaires, each of them and their responses will be individually analyzed to assess the effectiveness of the experience based on the different considered factors. The various measures of factors to be collected by the questions and the main characteristics of the chosen questionnaires are summarized in Table~\ref{tab:questionnaire}.

\begin{table}
\caption{Measures of the effectiveness of the VR experience for the user}\label{tab:questionnaire}
\begin{tabular}{p{2cm} p{4cm} p{2.15cm} p{2.5cm}}
\hline
\textbf{Measures} & \textbf{Definition}               & \textbf{Adapted from}               & 
\textbf{Questionnaire features} \\ \hline
Features of users     & Different features of the participants, including: Gender, Age and number of previous VR experiences  & \multicolumn{1}{c}{-} & 3 items. Multi-choice questions                \\
Sickness              & Side effects caused by the application that may affect the experience                           & VR Sickness questionnaire (VRSQ)~\cite{Kim2018}    & 9 items. 4-point Likert scale (from Not at all = 0, Very much = 3)              \\
Presence              & Comprises assessing the extent to which users perceive themselves to be genuinely immersed within the virtual environment                       & Slater-Usoh-Steed (SUS) questionnaire~\cite{Usoh2000}  & 6 items. 7-point Likert scale (from Not at all = 1 to Very much = 7)                 \\
Empathy improvement         & Self-reported considered improvement in empathy towards dyslexic students                       & Toronto Empathy Questionnaire (TEQ)~\cite{Spreng2009}      & 16 items. 5-point Likert scale (from Never = 0 to Always = 4)               \\

\end{tabular}
\end{table}

\subsection{{Procedure}}

{The evaluation was conducted in two stages. First, a preliminary test with 32 non-dyslexic participants was performed using an early version of the game to assess its usability and detect potential adverse effects, measured via the VRSQ (see Section~\ref{sec:sickness}). Based on the results, the experience was refined and validated for broader use.}

{In the final study, 101 participants completed the TEQ empathy questionnaire before using the application. After a four-month interval, they played the VR experience and answered the TEQ again to assess long-term changes. Additionally, the SUS and VRSQ were administered immediately after the session, and qualitative feedback was gathered to complement the quantitative results.}

\subsection{{Preliminary testing and prototype evaluation}}\label{sec:sickness}

\begin{figure}[ht]
        \centering
        \includegraphics[width=0.95\textwidth]{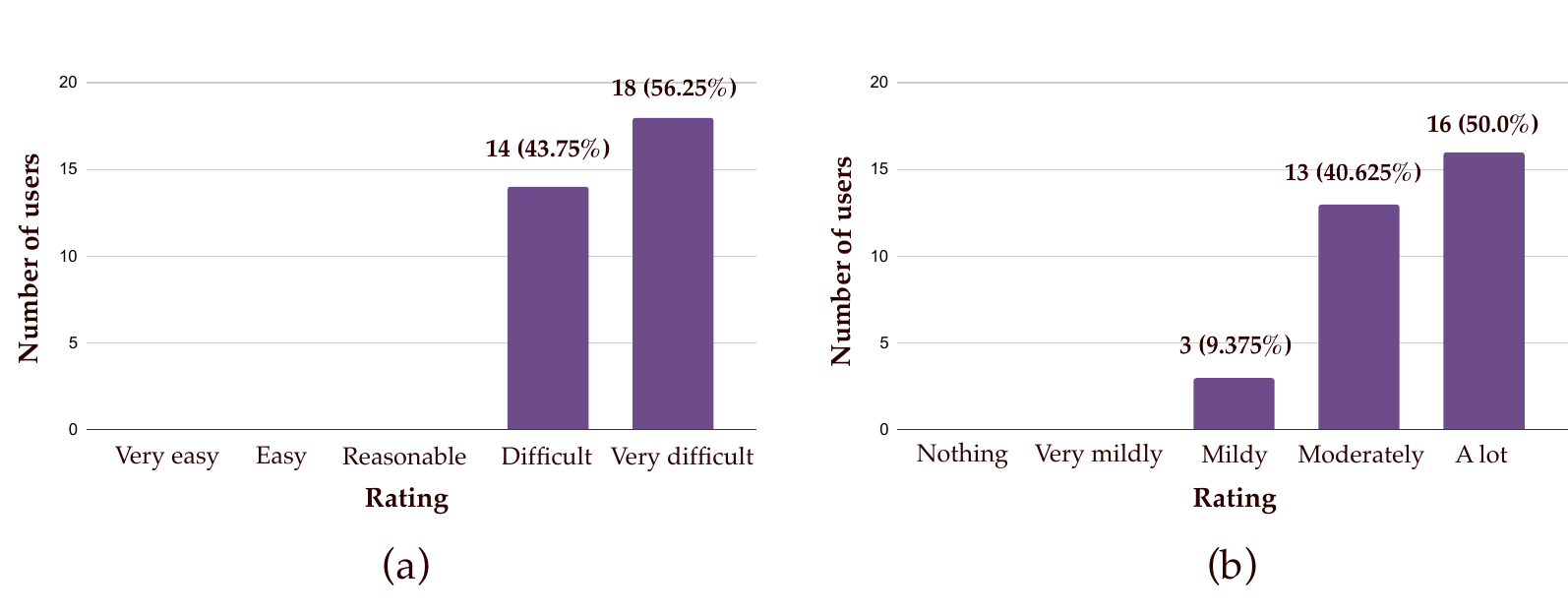}
        \caption{Results obtained after testing ``The Magic Potion'' in its initial version: (a)~``How challenging did you find the proposed task?'', and (b)~``How much did your empathy level increase after the VR experience?''}
        \label{fig:surv}
    \end{figure}

A first version of the game, available only in hard mode and in English, was tested by 32 non-dyslexic participants. The majority were unable to complete the task unassisted, with only 10 managing to finish the entire game. These individuals also reported experiencing frustration due to the inability to read the words correctly, noting that this frustration tended to decrease as compensatory tools were incorporated. Figure~\ref{fig:surv} shows two of the main questions {presented} to the participants after playing the game, the first referring to the difficulty of performing the proposed task and the second to the growth of their empathy towards people with dyslexia, both on a 5-point Likert scale. From this, we were able to conclude that awareness towards phonological dyslexia and towards the importance of support tools increased, thus fulfilling the main objective of the experience. However, due to the absence of rigorous validation of the game’s utility and the need to address certain errors identified during this initial phase, a more comprehensive data collection and analysis were required.

In addition, this preliminary study was used to assess the feasibility of the application's use with regard to potential symptoms that exposure to a virtual environment might produce during the experience.
For this purpose, the participants were subjected to the VRSQ. 
{The results showed an average score of 0.24 for the oculomotor component and 0.41 for disorientation (on a scale from 0 to 3), which is sufficiently low to ensure the viability of using the application to foster empathy without being adversely affected by other side effects typical of VR.}

\section{{Quantitative }results }\label{Experiments}

{This section describes the quantitative evaluation of ``The Magic Potion'' using validated questionnaires that assess key aspects addressed by the game, such as empathy improvement and sense of presence. The results are then analyzed and discussed to evaluate the effectiveness of the experience.}

\subsection{Measuring presence}
To calculate the relative impact of the developed virtual environment, the participants were subjected to the SUS questionnaire~\cite{Usoh2000}. After participating in the experience, the participants were asked to answer its six questions{, each rated on a 7-point Likert scale.} Obtained results align with {those reported in previous studies using the SUS questionnaire. For instance,~\cite{Usoh2000} and~\cite{Birrenbach2023} reported mean SUS scores of 3.8 and 4.4, respectively, in other VR simulations. In our study, participants scored a mean of 3.65 ± 0.95, and an interquartile range (IQR) of [2.96–4.42] (compared to [3.8–5.5] in~\cite{Birrenbach2023}, where the sense of presence and realism played a more central role). Additionally, the average number of responses rated 6 or 7 on the scale (SUS Count) was 1.25 ± 0.91, which is comparable to the 1.0 ± 1.7 reported for virtual training in~\cite{Usoh2000}. These slightly lower scores are expected, as the environment was intentionally designed to prioritize usability and comfort over photorealism, ensuring participants could carry out the task effectively while remaining engaged and immersed.}

\subsection{Empathy improvement}

\begin{figure}[ht]
    \centering
    \includegraphics[width=0.7\textwidth]{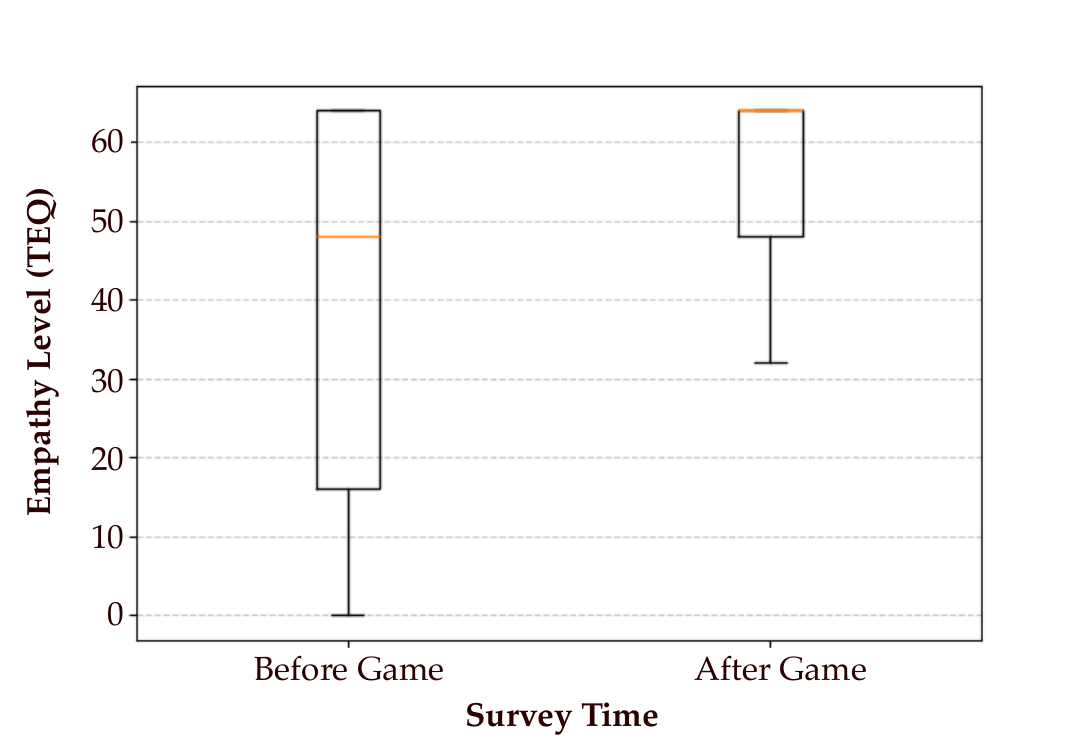}
    \caption{{Boxplot reflecting the TEQ scores before and after the VR experience.}}
    \label{fig:boxplotTEQ}
\end{figure}

Since the objective is to calculate the empathy enhancement with the use of the proposed VR serious game, the TEQ was first performed globally on 112 non-dyslexic university students and professors who had not experienced the developed VR serious game. From them, an average TEQ score of 39.45 was obtained. Four months later, TEQ was also administered to the 101 users who {actually participated in the serious game}. From them, the average TEQ score obtained was 58.77. {Figure~\ref{fig:boxplotTEQ} shows a boxplot comparing TEQ scores before and after the game. Before the experience, scores show a wider spread, with a substantial number of participants showing low empathy levels. However, after participating in the game, the distribution becomes narrower and shifts upward, with most scores concentrated near the upper side of the scale. This indicates that after participating in the VR experience a more homogeneous and higher empathetic response among participants was obtained.} 

The score obtained for people who had not participated in the experience is a little lower than that normally obtained in other empathy studies{, suggesting limited baseline awareness regarding students with dyslexia.}

{Although there are no established benchmarks specifically for empathy toward individuals with dyslexia, several general empathy studies using the TEQ in adult populations can be useful as reference values. In previous works using the TEQ, average scores above 45 are generally considered indicative of good empathy levels in adult populations. For example,~\cite{Baig2023} reported a mean TEQ score of 46.84 evaluating 347 medical students;~\cite{Bin2022} obtained an average score of 42.31 for 941 Saudi people, also medical students; and \cite{Totan2012} found an average of 38.37 assessing the empathy of 698 Turkish university students. Compared to this literature, the baseline score in our study (39.45) reflects lower-than-average empathy towards dyslexics, while the post-experience score (58.77) clearly exceeds the expected threshold.}

{Finally,} it can be determined that the increase in empathy in terms of the TEQ is 19.32 points, representing approximately a 20\% increase compared to the empathy calculated for these people in the university community before experiencing ``The Magic Potion'' four months earlier. {A Wilcoxon signed-rank test was conducted to assess whether the observed increase in TEQ scores was statistically significant. The results confirmed a significant difference between pre- and post-experience TEQ scores (W = 32.0, p < 0.001), supporting the effectiveness of the VR experience in enhancing empathy.}

\section{{Qualitative results}}
Once a quantitative analysis based on different metrics and questionnaires from the literature has been performed, it is also important to consider other aspects and observations that cannot be measured in this way. For this purpose, in this section, the opinions given by the participants after completing the experience of ``The Magic Potion'' are analyzed, as well as the most common behaviors identified by external observers. Additionally, the game is compared with other applications designed to support students with dyslexia, highlighting the innovations it offers. Similarly, during the comparison, the main limitations found in this type of VR tool, and more specifically in ``The Magic Potion,'' are also indicated.

\subsection{Participants' feedback and common behaviors}\label{sec:feedback}
{Participants provided qualitative feedback and exhibited common behavioral patterns throughout the experience. The most interesting and recurring observations included:}

\begin{itemize}
    \item \textbf{Difficulty in adapting to the game's movement and controls:} This was particularly indicated by some users who did not progress beyond the first level because they were unable to adapt to the mobility within the virtual environment. In contrast, those who completed the game praised the utility of the various locomotion systems implemented.
    \item \textbf{Issues using the audio guide:} Several users reported being unable to properly activate the audio guide function during the third phase (touching the name of the ingredients is the way to do it). These users were typically the same individuals who experienced difficulties in adapting to movement within the virtual environment, likely due to their lack of familiarity with such kind of systems. However, overall, the audio guide was deemed the most useful tool for completing the task.
    \item \textbf{Discomfort feeling during the experience:} Three-quarters of the participants reported feelings of fear, loneliness, dejection, and helplessness due to the features of the challenge and the environment. The teacher's reprimands, as well as Sam's deteriorating condition, also created feelings of frustration and powerlessness for such users. This effectively fulfilled the intended role of these NPCs within the game.
    \item \textbf{Significant challenge for the players:} Players emphasized the considerable difficulty of the challenge, especially during the first two levels of the game before obtaining the audio guide. In this way, it is demonstrated that the game achieves its objective of illustrating the necessity of compensatory tools for individuals with dyslexia.
    \item \textbf{Increase in awareness towards phonological dyslexia:} The majority of participants affirmed that they had never imagined that reading difficulties could have such an impact on a simple task, indicating an increase in empathy towards individuals with dyslexia.
    \item \textbf{Shift in perspective regarding compensatory tools:} A minority of users thought before playing the game that giving compensatory tools was unfair with respect to their non-dyslexic classmates. However, after completing the experience, all of them indicated a better understanding of the importance of these tools in facilitating the educational journey of individuals with dyslexia, thereby fulfilling the objectives of this work.
\end{itemize}

Concerning common behaviors, the following ones have been deemed noteworthy:

\begin{itemize}
    \item The majority of users adhered to the implicit game sequence, initiating the timer before starting the task. However, a small number of them, specifically 12 over 101, took advantage of available information before starting the first level. They searched for some of the necessary ingredients and arranged them on the table before activating the timer, a strategy aimed at saving time for task completion.
    \item Due to the human need to avoid obstacles and disadvantageous situations, a significant portion of the users requested the possibility to switch the text to the ``normal'' alphabet after five minutes of the experience, but upon being informed that it was not possible, they attempted to continue with the challenge.
    \item The hard game mode was attempted and successfully completed in the third stage by only 10 of the users who reached this stage in the easy mode, granting them a ten-minute break. The rest of the users declined the proposal due to fatigue and the difficulty they had encountered even with the easy mode.
\end{itemize}

Much of the information gathered about these opinions and behaviors aligns with the anticipated outcomes of the experience design. This includes the sentiment of frustration, leading some participants to disengage, and the increase in empathy towards individuals with phonological dyslexia. Additionally, encountered issues, such as the difficulty of adapting to the controls and movement within the game, will be considered for resolution in future experiences.

\subsection{{Overview of existing} dyslexia applications}

{To provide context and illustrate the positioning of our work, a comparative review of existing applications focused on dyslexia has been conducted.}
The applications for comparison are: ``The Magic Potion,'' ``Out of the Box''~\citep{yeguas-bolivar_determining_2022}, ``Dytective''~\citep{Rello2016}, ``Cosmic Sounds''~\citep{Brennan2022} and ``FORDYS-VAR''~\citep{rodriguez_cano_tecnologias_2021}. Comparison is summarized in Table~\ref{tab:comparison}, and has been conducted on the basis of the following elements:

\begin{itemize}
    \item \textbf{Reinforce reading (Reading):} The application provides a form of reinforcement to enhance reading for users with dyslexia.
    \item \textbf{Dyslexia detection (Detection):} The application aims to predict if the user has any type and degree of dyslexia.
    \item \textbf{Mobile:} Indicates whether the application can be used on mobile devices or requires any special device.
    \item \textbf{Kids:} The application is intended for children under 16 years old. Meta Quest 2 devices are not recommended for children under 13 years old. Therefore, if they wish to try applications deployed on these devices, they should do so under adult supervision.
    \item \textbf{Adults:} The application is intended for an adult audience.
    \item \textbf{Dizziness:} The application may cause dizziness in sensitive users.
    \item \textbf{Empathy}: The application aims to increase awareness and empathy towards individuals with dyslexia.
\end{itemize}

\setlength{\tabcolsep}{2pt}

\begin{table}[ht]
\caption{Qualitative comparison of different applications developed to support individuals with dyslexia.}\label{tab:comparison}
\begin{tabular}{cccccccc}
\hline
\textbf{Application} & \textbf{Reading} & \textbf{Detection} & \textbf{Mobile} & \textbf{Kids} & \textbf{Adults} & \textbf{Dizziness} & \textbf{Empathy} \\ \hline
The Magic Potion     & X                           & X                         & X               & \checkmark     & \checkmark       & \checkmark          & \checkmark        \\ \hline
Out of the Box       & X                           & \checkmark                 & \checkmark       & \checkmark     & \checkmark       & \checkmark                  & X                \\ \hline
Dytective            & \checkmark                   & \checkmark                 & \checkmark       & \checkmark     & X               & X                  & X                \\ \hline
Cosmic Sounds            & \checkmark                   & X                 & \checkmark       & \checkmark     & X               & X                  & X                \\ \hline
FordysVAR            & \checkmark                   & X                         & X       & \checkmark     & X               & \checkmark          & X                \\ \hline
\end{tabular}
\end{table}

\setlength{\tabcolsep}{10pt}

{This comparison clearly indicates that, although all reviewed applications are designed as support tools for individuals with dyslexia, they pursue very different objectives. Most of them are focused on enhancing reading skills in school-aged children, and only “Out of the Box” and our proposal are intended for adult users. In contrast to training tools, our application, ``The Magic Potion'', addresses dyslexia from a different perspective: by promoting empathy among individuals in the educational environment, particularly in university and higher education settings. This underexplored focus highlights the originality and relevance of our approach, offering a complementary and awareness-raising tool rather than a direct intervention.}

\section{{Discussion and limitations}}

{The results obtained through the TEQ demonstrate a significant increase in empathy levels following the VR experience ``The Magic Potion''. Participants showed an average gain of 19.32 points (approximately a 20\% increase), confirming a consistent trend across users. This indicates that VR activity can effectively foster greater empathetic understanding toward individuals with dyslexia. Qualitative feedback collected after the sessions supports these quantitative findings. Many participants reported increased awareness of the frustration and confusion dyslexic individuals may experience in academic contexts. The emotional impact of the experience was frequently highlighted, with several users suggesting the usefulness of such tools in teacher training programs or inclusion workshops.}

{On the other hand, in the broader landscape of applications related to dyslexia, most applications focus on skill development or diagnostic support for children. Very few target adult users or seek to influence their perceptions of dyslexia. In contrast, ``The Magic Potion'' contributes a novel approach, instead of aiming to improve reading abilities, it aims to promote empathy towards dyslexics within the university context. This distinction positions our application as an awareness-oriented tool, adding relevance to current efforts in inclusive education.}

{However, several limitations must be considered. From a technological point of view, the use of VR comes with accessibility challenges. Devices like the Meta Quest 2~\& 3, while offering the level of realism needed for this type of immersive simulation, are not as widely available or affordable as mobile or web-based alternatives. Moreover, some users, particularly older individuals or those unfamiliar with video games, may experience discomfort or dizziness~\citep{Saredakis_2020}.}

{In addition to these general barriers associated with VR, other specific limitations of our deployment were observed. One of the main issues was the time required to complete the full experience, which lasted approximately twenty minutes. This made it difficult for many participants to try the experience during short sessions, and some chose to leave due to time constraints. Another constraint was related to physical space. Users enjoyed the experience more in wide, open areas, especially those unfamiliar with movement via controllers. Since Meta Quest 2 \& 3 headsets require indoor usage conditions for optimal performance and maintenance~\citep{Meta}, securing large indoor spaces for extended sessions posed logistical challenges.}

\section{Conclusion}\label{Conclusion}
The objective of this work was to implement and evaluate the utility of a serious game in VR aimed at increasing awareness and empathy towards students with phonological dyslexia. The experience has been developed and evaluated in stages, during different test phases until reaching its final version, which has been tested by 101 non-dyslexic individuals. All participants were asked to complete a series of questionnaires after the experience to validate the tool's usefulness. Those who participated in the initial version of the game were given the VRSQ to validate the game's viability in terms of health side effects caused by players' exposure to the virtual environment. After this, with the final version of ``The Magic Potion'', 101 non-dyslexic people participated in the experience. They were given the SUS to measure the quality of the experience in terms of the sense of presence felt by the players in the virtual environment, yielding good results. The increase in empathy was measured using the TEQ, which was first administered to 112 non-dyslexic people who had not participated in the experience and then to the 101 participants who did play the game. The results showed an average increase in empathy of 20\%, validating the tool's usefulness as a means of promoting empathy towards university students with dyslexia. In addition, qualitative opinions reported by the users as well as the observations of their behavior, made during the development of the experience, confirmed the quantitative results and the conclusions arising from them. 

In summary, the study revealed a promising increase in empathy and awareness towards dyslexia among the participants following the gaming experience. The results demonstrate that the implementation of VR as an educational tool could be effective in fostering understanding and support for those facing learning difficulties such as dyslexia. Furthermore, experimentation with different levels of difficulty and the inclusion of different compensatory tools as participants failed in the game has enabled them to understand the necessity of such kind of aids for dyslexic students when performing tasks or exams.

This work will be enriched by designing new VR games, focused on the types of dyslexia other than phonological dyslexia. Addressing all types of dyslexia should increase awareness of the specific learning disorder and foster empathy toward those affected, as well as highlight the significant differences in the symptoms they experience. These endeavors aim to foster the inclusion of diverse student profiles facing educational challenges, with the overarching goal of achieving the highest grade of inclusive education.

\bmhead{Acknowledgements}

José Manuel Alcalde Llergo is a PhD student enrolled in the National PhD in Artificial Intelligence, XXXVIII cycle, course on Health and life sciences, organized by Università Campus Bio-Medico di Roma. He is also pursuing his doctorate in co-supervision at the Universidad de Córdoba (Spain), enrolled in its PhD program in Computation, Energy and Plasmas.

\bmhead{Funding}
These results are framed in “VRAIlexia –
Partnering Outside the Box: Digital and Artificial Intelligence
Integrated Tools to Support Higher Education Students with
Dyslexia” funded by the Erasmus+ Programme2014-2020 –
Key Action 2: Strategic Partnership Projects. AGREEMENT
n. 2020-1-IT02-KA203-080006.This article has been funded
with support from the European Commission.

\bmhead{Data availability}

Data underlying the results presented in this paper are available from the corresponding author on reasonable request.

\section*{Statements and Declarations}
\textbf{Conflict of interest} The authors have no competing interests to declare that are relevant to the content of this article.

\bibliography{InTheShoes}

\end{document}